\documentclass{revtex4}

\usepackage{graphicx}
\usepackage{dcolumn}
\usepackage{amsmath}
\usepackage{epsfig}
\usepackage{amssymb}

\newcommand{\ket}[1]{\left\vert#1\right\rangle}
\newcommand{\bra}[1]{\left\langle#1\right\vert}

\newcommand{\dmat}[2]{\ket{#1}\!\!\bra{#2}}
\newcommand{\half}{\frac{1}{2}}

\newcommand{\beq}{\begin{equation}}
\newcommand{\eeq}{\end{equation}}
\newcommand{\bea}{\begin{eqnarray}}
\newcommand{\eea}{\end{eqnarray}}

\newcommand{\tr}{\mbox{Tr}}

\makeatletter
\def\btt#1{\texttt{\@backslashchar#1}}
\DeclareRobustCommand\bblash{\btt{\@backslashchar}}
\makeatother
\def\Bid{{\mathchoice {\rm {1\mskip-4.5mu l}} {\rm
{1\mskip-4.5mu l}} {\rm {1\mskip-3.8mu l}} {\rm {1\mskip-4.3mu l}}}}

\newcommand{\la}{\lambda}

\newcommand{\sigmav}{{\vec \sigma}}

\newcommand{\tv}{\vec{t}}
\newcommand{\lav}{{\vec \lambda}}
\newcommand{\nv}{{\vec n}}

\newcommand{\vv}{{\vec v}}

\begin{document}

\title{Affine Maps of the Polarization Vector for Quantum Systems of
  Arbitrary Dimension}
\author{Mark S. Byrd$^{1,2}$, C. Allen Bishop$^1$, Yong-Cheng Ou$^1$}
\affiliation{$^1$Physics Department, $^2$Computer Science Department, Southern Illinois University, 
Carbondale, Illinois 62901}

\date{\today}

\begin{abstract}
The operator-sum decomposition (OS) of a mapping from one density 
matrix to another has many applications in quantum information 
science.  To this mapping there corresponds an affine map which 
provides a geometric description of the density matrix in terms 
of the polarization vector representation.  This has been thoroughly 
explored for qubits since the components of the polarization vector
are measurable quantities (corresponding to expectation values of
Hermitian operators) and also because it enables the description of map
domains geometrically.  Here we extend the
OS-affine map correspondence to qudits, briefly discuss general 
properties of the map, the form for particular important cases, 
and provide several explicit results for qutrit maps.  
We use the affine map and a singular-value-like decomposition, 
to find positivity
constraints that provide a symmetry for small polarization vector
magnitudes (states which are closer to the maximally mixed state)
which is broken as the polarization vector increases in magnitude
(a state becomes more pure).
The dependence of this symmetry on the magnitude of the polarization
vector implies the polar decomposition of the map can not be used 
as it can for the qubit case.  However, it still leads us to a
connection between positivity and purity for general d-state systems.
\end{abstract}

\maketitle

\tableofcontents


\section{Introduction}

The study of maps of density operators has become an active
field of research in quantum information theory.  Such maps 
are used to describe open-system evolution, i.e., evolution of quantum
systems which interact with their environment in a non-trivial way.  
Such maps are also useful for describing methods to alleviate such
noise and, in some cases, as indicators of entanglement 
\cite{Peres,Horodeckis}.  These are especially important in the study
of quantum information processing.  

General maps of density operators to density operators were first studied by
Sudarshan, Mathews and Rau (SMR) \cite{Sudarshan:61}.  They provided a method
for describing open-system evolution and referred to such maps as
``dynamical maps.''  This name is particularly appropriate for
describing noise since detailed properties of the environment causing
the noise are often unknown.  Often these maps are
written in what is called an operator-sum representation (OSR). 
Kraus later considered completely positive maps \cite{Kraus:83}.
The assumption of complete positivity is sometimes useful but is not a 
necessary assumption for open system evolution. The physical 
implications of the complete positivity assumption has recently been
discussed at length in the literature 
\cite{Pechukas:94,Pechukas+Alicki:95,Jordan:04,Shaji/Sudarshan:05,Shabani/Lidar:09a,Rodriguez/etal:10}.
Such work is not only important for quantum error correction and the
description of noise in quantum systems \cite{Shabani/Lidar:09b}, but
also for quantum control more generally.  (See 
Ref.~\cite{Dong/Petersen:09,Brif/etal:10} for recent
reviews of quantum control theory and applications.)  

Studies of maps of qubits have been extensive.  Some notable 
discussions are found in Refs.~\cite{Nielsen/Chuang:book,King/Ruskai}
where they describe a geometric picture of the spaces of states of 
qubits and
the ranges of maps of qubits.  This is done using the Bloch vector, 
or polarization vector parameterization of the density operator 
\cite{Mahler:book,Jakob:01,Arvind:97} and an
affine mapping of the polarization vector.  This affine map is
essentially equivalent to the OSR, but is more geometrical, leading to
geometric constructions which are sometimes helpful tools for
visualization as well as analysis.  

The subject of noise in quantum systems governs most
of the discussion of such maps in Ref.~\cite{Nielsen/Chuang:book}, and
this will also be the case here.  The discussions in
Refs.~\cite{Nielsen/Chuang:book,King/Ruskai} are based on the Bloch
vector parameterization of the density operator.  The generalization
of the 
Bloch vector is known as the generalized Bloch vector, coherence
vector, or polarization vector.  Since the dimension of this vector 
grows with the dimension of the Hilbert space as $d^2-1$, where $d$ is
the dimension of the Hilbert space, the simple three dimensional
picture (for $d=2$) becomes more difficult to use and one loses the
ability to visualize the entire space.  However, 
the polarization vector parameterization 
has several appealing properties.  (1) The components
of the polarization vector are measurable quantities; they are
proportional to the expectation values of Hermitian operators.  (2) The
trace condition (the trace of the density operator must be one) and
the hermiticity of the density operator become apparent in this picture.
(3) The positivity of the density operator can be expressed in terms of
the components of the density operator and the magnitude of the
polarization vector is directly related to its purity 
(\cite{Byrd/Khaneja:03,Kimura}).  (4) The 
Casimir invariants, quantities which are invariant under all unitary
operators, are easily written down in terms of the polarization
vector \cite{Byrd/Khaneja:03}.  (5) A tensor product basis may be 
used so that subsystems 
of interest are manifest \cite{Jakob:01,Byrd/Khaneja:03}.  
Clearly these are not entirely independent
properties.  However, they are useful properties.  For these reasons,
this picture, which generalizes the Bloch equations, is becoming more
widely used in both theory and experiment.  
This begs the question, which we attempt to answer, how
can we fully utilize this picture?

In this paper we discuss the generalization of affine maps of the
polarization vector to systems with $d$ dimensions.  We provide 
explicit expressions for the affine map of the polarization vector in 
terms of the OSR elements.  (The representation of a map of the
polarization vector generally contains linear and translational
terms.)  Our explicit calculation 
provides a direct link between the OSR, 
pioneered by Sudarshan, Mathews and Rau \cite{Sudarshan:61}
and the affine map 
picture as is done in Refs.~\cite{Nielsen/Chuang:book,King/Ruskai} for 
qubits.  We discuss the possibility of extending the description of
the positivity domain for qubits \cite{King/Ruskai} to 
$d$-dimensional systems and find a 
continuous symmetry breaking description which describes the inability 
of one to apply a simple singular value decomposition (SVD) to all
density operators.  This symmetry breaking provides a picture which 
compliments the work of Kimura and Kossakowski describing the
positivity domains for density 
operators of $d$-state systems \cite{Kimura/Kossakowski} but in
a very different way.  

Specifically, in Section \ref{sec:bkgd}, a brief review of the
derivation of the general operator-sum decomposition (OSR) from the
dynamical map is given, along with a brief review of the polarization
vector parameterization of the density operator.  In Section
\ref{sec:Affinemap} the form of the affine map is derived from the OSR
and some properties of the affine map are given.  In Section
\ref{sec:examples} physical examples of noise are given to show how
the polarization vector changes for some important types of noise.  
We then examine a particularly important basis for the operators
comprising the OSR in Section \ref{sec:unandortho}.  Having
established the relation between the OSR and affine map, we show how
one can find the affine map directly from the corresponding dynamical
map in Section \ref{sec:dyntoOSR}.  Section \ref{sec:svd} contains a
review of the singular value decomposition of the affine map for
qubits and our attempted generalization.  Finally, concluding remarks
and future directions are provided in Section \ref{sec:concl}.


\section{Background}
\label{sec:bkgd}

In this section we provide background for what follows.  This primarily 
follows the results of Ref.~\cite{Sudarshan:61}.


\subsection{Dynamical Maps and the SMR Decomposition}
\label{sec:osr}

As did Sudarshan, Mathews, and Rau, let us
consider a general mapping from one Hermitian matrix to another 
of the form
\begin{equation}
\label{eq:Asnosub}
\rho^\prime = A \rho,
\end{equation}
or more explicitly
\begin{equation}
\label{eq:As1}
\rho^\prime_{r^\prime s^\prime} = A_{r^\prime s^\prime,rs} \rho_{rs}.
\end{equation}
Since the density matrix is
required to be Hermitian $\rho = \rho^\dagger$, positive
semi-definite, $\rho \geq 0$, and have
trace one  $\mbox{Tr}\rho = 1$ the mapping 
$A$ must have the following properties in order for it to map 
density operators to density operators:
\begin{equation}
\label{eq:Aherm}
A_{sr,s^\prime r^\prime} = (A_{rs,r^\prime s^\prime})^*, \;\;
\end{equation}
which ensures hermiticity, 
\begin{equation} 
\label{eq:postr}
x_r^*x_sA_{sr,s^\prime r^\prime}y_{s^\prime}y_{r^\prime} \geq 0, \;\;
A_{rr,s^\prime r^\prime} = \delta_{s^\prime r^\prime},
\end{equation}
which ensure positivity and the trace condition respectively.  
SMR then introduced a matrix $B$, related to
$A$ by relabeling, such that
\begin{equation}
\label{eq:B}
B_{rr^\prime,s s^\prime}\equiv A_{sr,s^\prime r^\prime}. 
\end{equation}
This matrix has the following properties:
\begin{equation}
\label{eq:Bherm}
B_{rr^\prime,s s^\prime}=(B_{ss^\prime,r r^\prime})^*, \;\;
\end{equation}
corresponding to Eq.~(\ref{eq:Aherm}), and 
\begin{equation}
z^*_{rr^\prime}B_{rr^\prime,s s^\prime}z_{ss^\prime} \geq 0, \;\;
B_{rr^\prime,r s^\prime} = \delta_{r^\prime s^\prime},
\end{equation}
which corresponds to Eqs.~(\ref{eq:postr}).  In this article we will 
allow the map to be more general initially and only require the first of 
these conditions corresponding to the hermiticity condition, 
Eq.~(\ref{eq:Aherm}) or, equivalently, Eq.~(\ref{eq:Bherm}).  
Since $B$ satisifies the hermiticity condition, it can be considered 
a Hermitian matrix and as such it is diagonalizable using a 
spectral decomposition, or eigenvector decomposition.  Letting
$\eta_k$ be the eigenvalues of $B$ the spectral decomposition
can be written as  
\begin{equation}
\label{eq:OSdecompcomponents}
B_{rr^\prime,s s^\prime} = \sum_k
               \eta_k
               \xi^{(k)}_{rr^\prime}\xi^{(k)\dagger}_{s^\prime s}.
\end{equation}
This is also sometimes written with indices supressed
as follows:
\begin{equation}
\label{eq:OSdecomp}
B = \sum_k \eta_k C_k^{\phantom{\dagger}} C_k^{\dagger},
\end{equation}
where $(C_k)_{rr^\prime}=\xi^{(k)}_{rr^\prime}$ are eigenvectors of
$B$ and $\eta_k$ its eigenvalues.    
This is shown in detail in \cite{Jordan:04}.  The $C_k$, as well as 
the density operator itself, may be written as a matrix or a vector.  

If the $\eta_k$ are all positive, they may be absorbed into the
$C$'s to arrive at the familiar form of the operator-sum decomposition:
\begin{equation}
\label{eq:dmatstsmr}
\rho^\prime = \sum_k
               A_{k}^{\phantom{\dagger}}\rho A_{k}^\dagger,
\end{equation}
where $A_k = \sqrt{\eta_k} C_k$ \cite{Krausnote}.


\subsection{Polarization Vector Representation of the Density Operator}

Before we present the polarization representation of density operators, 
it is important to provide our conventions.  These are contained in
Ref.~\cite{Byrd/Khaneja:03} which follow Refs.~\cite{Arvind:97,Byrd:98}.  
However, they are not completely standard; 
see for example \cite{Kimura,Mahler:book,Jakob:01} 
for other conventions.  A 
density operator on a $d$-dimensional Hilbert space ${\cal H}_d$ 
will be represented using a set of traceless, Hermitan 
matrices $\{\lambda_i\}$, $i=1,2,...,d^2-1$ with the normalization 
condition $\tr(\lambda_i\lambda_j)=2\delta_{ij}$, commutation
$[\lambda_i,\lambda_j] = 2if_{ijk}\lambda_k$ ($f_{ijk}$ are the 
totally antisymmetric structure constants),  
and anticommutation relations 
$\{\lambda_i,\lambda_j\} = \frac{4}{d}\Bid \delta_{ij} + 2d_{ijk}\lambda_k$ 
(the $d_{ijk}$ are the totally symmetric $d$-tensor components), 
where the sum over repeated indices is to be understood unless
otherwise stated.  (In some cases the sum is displayed explicitly for
emphasis.)  
These three relations may be summarized using the following equation
\begin{equation}
\lambda_i\lambda_j = \frac{2}{d}\Bid \delta_{ij} 
                     + d_{ijk}\lambda_k + if_{ijk}\lambda_k.
\end{equation}
The density operator can now be written as 
\begin{equation}
\rho = \frac{1}{d}\left(\Bid + b \nv \cdot \lav\right),
\end{equation}
where $b = \sqrt{(d(d-1)/2)}$.  The ``dot'' product is a sum 
over repeated indices,
\begin{equation}
\vec{a}\cdot \vec{b} = a_ib_i = \sum_{i=1}^{d^2-1}a_ib_i.  
\end{equation}
Any complete set of $d^2-1$ mutually trace-orthogonal, Hermitian matrices 
can serve as a basis and can be chosen to satisfy 
the conditions given here.  The components of the polarization vector
$\nv$ are proportional to the expectation values of the set of
Hermitian observables $\lav$:
\begin{equation}
n_i = \frac{d}{2b}\tr(\lambda_i\rho). 
\end{equation}
Therefore these are directly measurable quantities which can be used
to completely specify any state, pure or mixed.

Pure states have the properties that 
\begin{equation}
\nv\cdot\nv = 1, \;\;\;\mbox{and} \;\;\; \nv\star\nv = \nv,
\end{equation}
where the ``star'' product is defined by 
\begin{equation}
\label{eq:defstar}
(\vec{a}\star\vec{b})_k = \frac{1}{d-2}\sqrt{\frac{d(d-1)}{2}}\;d_{ijk}a_ib_j.
\end{equation}
For later use, a ``cross'' product between two coherence vectors can 
also be defined by 
\begin{equation}
\label{eq:defcross}
(\vec{a}\times\vec{b})_k = f_{ijk}a_ib_j.
\end{equation}
The set of mixed states can be specified in terms of a set of
positivity conditions \cite{Byrd/Khaneja:03,Kimura}.


\section{Affine Maps from the OSR}

\label{sec:Affinemap}

In this section we obtain the affine map of the polarization vector in
terms of the components of the OSR 
of the dynamical map.  However, we begin with a 
general case, as is done in Ref.~\cite{King/Ruskai}, and restrict to
particular classes of maps which are often physically relevant.
Furthermore, we do not restrict to completely positive maps.  
The affine map provides a final connection 
between these three different forms of the map: the dynamical map $A$
(or equivalently $B$), the OSR, and the affine map.


\subsection{Explicit form for the Affine Map}

Let the operator-sum decomposition for a map from one density 
operator to another be given by
\begin{equation}
\label{eq:map}
\rho^\prime = \Phi(\rho) = \sum_k \eta_k C_k \rho C^\dagger_k.
\end{equation}
Following Ref.~\cite{King/Ruskai} which 
treats the qubit case, 
the $C_k$ can be represented by (complex) linear combinations 
of the $\{\lambda_i\}$, as 
\begin{equation}
\label{eq:compparC}
C_k = v_{0k}\Bid +\vec{v}_k\cdot \vec{\lambda},
\end{equation}
where $v_{0k}$, $v_{ik} \in \mathbb{C}$.  

The following identity will be used repeatedly in our derivations
\begin{equation}
\label{eq:vecid}
(a\Bid + \vec{u}\cdot\lav)(b\Bid + \vec{w}\cdot\lav) = 
           \left(ab +\frac{2}{d}\vec{u}\cdot\vec{w}\right)\Bid 
             + \left(a\vec{w} + b\vec{u} + i\vec{u}\times\vec{w} 
                 + \frac{1}{c}\vec{u}\star\vec{w}\right)\cdot \lav,
\end{equation}
where the definitions Eqs.~(\ref{eq:defstar}) and (\ref{eq:defcross}) were 
used along with the following definition, 
\begin{equation}
c = \sqrt{\frac{d(d-1)}{2}}\;\frac{1}{d-2}.
\end{equation}

Two special classes of maps are particularly important, unital maps
and trace preserving maps.  They are defined by 
\begin{enumerate}
\item $\Phi$ is called {\it unital} if $\Phi(\Bid) = \Bid$, 
\item $\Phi$ is called {\it trace preserving} if $\forall \; \rho$, 
$\tr(\Phi(\rho))=\tr(\rho)$.
\end{enumerate}
If the map is unital, they by explicit calculation using 
Eq.~(\ref{eq:map}),  
\begin{equation}
\label{eq:u1}
\sum_k \eta_k \left[|v_{0k}|^2 + \frac{2}{d}\vec{v}_k\cdot\vv^*_k\right] =1,
\end{equation}
and 
\begin{equation}
\label{eq:u2}
\sum_k \eta_k  \left(v_{0k}\vv^*_k + v_{0k}^*\vv_k + i\vv_k\times\vv_k^* 
         +\frac{1}{c}\vv_k\star\vv_k^*\right) = 0.
\end{equation}

Similarly, if the map is trace preserving, then
an explicit calculation shows that
\begin{equation}
\label{eq:tp1}
\sum_k \eta_k \Big[|v_{0k}|^2 + \frac{2}{d}\vec{v}_k\cdot\vv^*_k\Big] =1,
\end{equation} 
and
\begin{equation}
\label{eq:tp2}
\sum_k \eta_k  \left(v_{0k}\vv^*_k + v_{0k}^*\vv_k + i\vv_k^*\times\vv_k 
         +\frac{1}{c}\vv_k^*\star\vv_k\right) = 0.
\end{equation}

Note that if the map is either unital {\it or} trace-preserving, then 
the condition given in Eq.~(\ref{eq:u1}) (equivalently 
Eq.~(\ref{eq:tp1})) holds.

To provide the explicit form of the map, we first use 
Eqs.~(\ref{eq:map}), (\ref{eq:compparC}), and 
(\ref{eq:vecid}), so that 
\begin{equation}
\Phi(\rho) = \sum_k \eta_k C_k\rho C_k^\dagger = \sum_k \eta_k (v_{0k}\Bid 
             + \vv_k\cdot\lav) \frac{1}{d}(\Bid 
              + b\nv \cdot \lav)(v_{0k}^*\Bid + \vv_k^*\cdot\lav), 
\end{equation}
can be rewritten and 
the identity and $\lav$ parts treated separately.  

First, consider the identity part.  
The coefficient of the identity can be written in the following 
form 
\begin{equation}
\sum_k\eta_k\left(\frac{1}{d}\Big[|v_{0k}|^2+\frac{2}{d}\vv_k\cdot\vv_k^*\Big] 
 + \frac{2b}{d^2}\left(v_{0k}^*\vv_k+v_{0k}\vv_k^*
        +i\vv_k^*\times\vv_k+\frac{1}{c}\vv_k^*\star\vv_k\right)\cdot\nv\right),
\end{equation}
If the map is unital, using Eqs.~(\ref{eq:u1}) 
and (\ref{eq:u2}), the coefficient of the identity reduces to 
\begin{equation}
\frac{1}{d}+\frac{4ib}{d^2}\sum_k\eta_k (\vv_k^*\times\vv_k)\cdot\nv.
\end{equation}
If the map is also trace-perserving, subtracting 
Eq.~(\ref{eq:u2}) from Eq.~(\ref{eq:tp2}) 
implies that the coefficient of the identity is $1/d$.

Now let us consider the non-identity part of the map.  
Denoting the result of the map by $\rho^\prime$, we have
\begin{equation}
\Phi(\rho) = \rho^\prime 
           = \frac{1}{d}\Big(\Bid + b\; \nv^\prime\cdot\lav\Big).
\end{equation}
Viewing the map as an affine map of the coherence vector $\nv$, 
\begin{equation}
\nv \mapsto \nv^\prime = T\nv +\vec{t},
\end{equation} 
so the components of $\vec{n}^\prime$ are given by 
\begin{equation}
n^\prime_q = T_{pq}n_p+t_q.
\end{equation}
Thus, after some rearrangement, 
the map can be written as
\begin{gather}
\label{eq:t}
\frac{1}{d}\sum_k\eta_k\left[v_{0k}^*\vv_k+v_{0k}\vv_k^*
        +i\vv_k\times\vv_k^*+\frac{1}{c}\vv_k^*\star\vv_k\right]\cdot\lav
\\
+\frac{1}{d}\sum_k\eta_k\bigg[ b|v_{0k}|^2\nv+\frac{2b}{d}(\vv_k\cdot\nv)\vv^*_k
          +ib(v_{0k}^*\vv_k\times\nv+v_{0k}\nv\times\vv_k^*)           
            +\frac{b}{c}(v_{0k}\nv\star\vv_k^*+v_{0k}^*\nv\star\vv_k )\nonumber \\
 \;\;\;\;\;\;\;\;+\frac{ib}{c}\big\{(\vv_k\star\nv)\times\vv_k^*
               +(\vv_k \times\nv)\star\vv_k^*\big\}
            -b(\vv_k\times\nv)\times\vv^*_k+\frac{b}{c^2}(\vv_k\star\nv)\star\vv_k^*
 \bigg] \cdot \lav,
\label{eq:To}
\end{gather}
where $T = \sum_k \eta_k  T_k$ and $\vec{t} = \sum_k \eta_k \vec{t}_k$.  
The $k^{\mbox{\scriptsize{th}}}$ term
(\ref{eq:t}), is 
\begin{equation}
\vec{t}_k = \frac{1}{b}\left[v_{0k}^*\vv_k+v_{0k}\vv_k^*
        +i\vv_k\times\vv_k^*+\frac{1}{c}\vv_k^*\star\vv_k\right].
\end{equation}
Note that $\vec{t}$ is zero for a unital map.  

The $k^{\mbox{\scriptsize{th}}}$ term (\ref{eq:To}) represents the action of the
{\it real} $(d^2-1)\times(d^2-1)$ matrix $T$.  
(The proof of the reality of $T$ is given below.)  The components of the linear 
part of the map, which we refer to as the $T$ matrix, may
then be written as
\begin{eqnarray}
T_{pq} &=& \frac{1}{d}\sum_k\eta_k\left[  \left(|v_{0k}|^2-\frac{2}{d}v_{r k}^*v_{r
    k}^{\phantom{*}}\right) \delta_{pq}
              +\frac{2}{d}(v_{p k}v_{q k}^* + v_{p k}^*v_{q k})
    +if_{rpq}(v_{0k}^*v_{r k}
    -v_{0k}v_{r k}^*)\right.
         \nonumber \\
    && +\; d_{rpq}(v_{0k}^*v_{r k} + v_{0k}v_{r k}^*)
        -id_{rpq}f_{srt}v_{t k}^* v_{s k}^{\phantom{*}}
         \nonumber \\
     &&
     +\left. (d_{qsr}d_{rtp}+d_{qtr}d_{spr}-d_{qpr}d_{str})v_{tk}^{\phantom{*}} v^*_{sk} \!\!\!\!\!\!\!\!\!\phantom{\sum_k}\right].
\label{eq:Tcomponents}
\end{eqnarray}

It is worth emphasizing that the $T$ matrix and the vector $\vec{t}$ 
specify an affine map of the polarization vector. We also note that the map 
is linear when 
$\vec{t} =0$.  This is equivalent to the dynamical 
map, $A$ (or $B$) specified in Section \ref{sec:bkgd} and also the 
operator-sum decomposition.  
Thus we have provided a mapping of a vector to a vector which corresponds 
to a general mapping of a density operator to a density operator.  The 
next question which we will answer in part is the following.  What are the 
properties of this map?  We are able to provide partial answers to this 
broad question for both the general map and for some particular cases of 
interest, which we do next.


\subsection{Properties of the Associated Linear Map}

In this subsection we show that the linear term in the affine map,
i.e., the $T$ matrix, is real.  Although this
must be true, it is shown explicitly since it 
is not obvious from the expression Eq.~(\ref{eq:Tcomponents}).
In addition, we provide general conditions for the matrix to be
symmetric.


\subsubsection{The $T$-matrix is real}

\label{sec:realitycheck}

It is not clear that the third to the last term in the 
curly brackets ($\{\}$), (\ref{eq:To}), which mixes star and 
cross products, is real.  If we can show that this is real, then we
will have shown that $T$ is real since the other terms are obviously
real.  Let us consider whether 
\begin{equation}
\label{eq:?realterms}
 +\frac{ib}{c}\sum_k\eta_k\big[\big((\vv_k\star\nv)\times\vv_k^*\big)\cdot\lav
                           +\big((\vv_k\times\nv)\star\vv_k^*\big)\cdot\lav\big]
\end{equation}
is equal to its complex conjugate.

To provide a sufficient condition, let us first consider a term with a
given $k$.  
From the identity, Eq.~(\ref{eq:jlid}):
\begin{equation}
\label{eq:j-lid}
d_{ijm}f_{mln}+d_{jlm}f_{min}+d_{lim}f_{mjn}=0
\end{equation}
we can form the following identities with any vectors 
$\vec{a},\vec{b},\vec{c}, \vec{\la}$.  (Here we will use $\vec{\la}$ 
as the basis for the Lie algebra, but in these identities, it could 
be any vector.)  First, we contract the LHS of Eq.~(\ref{eq:j-lid}) 
with $a_ib_jc_l\la_n$ and find:  
\begin{equation}
[(\vec{a}\star\vec{b})\times\vec{c}]\cdot \vec{\la} 
               + [(\vec{b}\star\vec{c})\times \vec{a}]\cdot\vec{\la} 
               + [(\vec{c}\star\vec{a})\times \vec{b}]\cdot\vec{\la} = 0. 
\end{equation}
For the next identity, we contract Eq.~(\ref{eq:j-lid}) with 
$a_jb_lc_n\la_i$ to obtain:
\begin{equation}
\label{eq:j-lvecid}
[(\vec{b}\times\vec{c})\star\vec{a}]\cdot\vec{\la} 
           + \vec{\la}\cdot[\vec{c}\times(\vec{a}\star\vec{b})]
	   + [(\vec{a}\times\vec{c})\star\vec{b}]\cdot\vec{\la} = 0.
\end{equation}
For our purposes, it is relevant to note the following symmetries in the 
indices.  The identity is invariant under the interchange of the following 
pairs of indices $(l,i)$, $(i,j)$, $(l,j)$, so that these are the only 
two different types of identities when we distinguish $\vec{\la}$ as 
a set of basis elements.  

We will use this second identity Eq.~(\ref{eq:j-lvecid}) to rewrite the 
first of the two terms in Eq.~(\ref{eq:?realterms}).  First, note that 
\begin{equation}
[\vv^*_k\times(\vv_k\star\nv)]\cdot \vec{\la} 
                    + [(\nv\times\vv^*_k)\star\vv_k]\cdot\vec{\la}
                    + [(\vv_k\times\vv^*_k)\star\nv]\cdot\vec{\la} = 0.
\end{equation}
Now the first term in this equation is the same (up to sign) as the 
first of the two terms in Eq.~(\ref{eq:?realterms}).  This allows us 
to rewrite Eq.~(\ref{eq:?realterms}) as
\begin{equation}
\frac{ib}{c}\left\{[(\nv\times\vv_k^*)\star\vv_k]\cdot \vec{\la} 
          + [(\vv_k\times\vv_k^*)\star\nv]\cdot\vec{\la}
          + [(\vv_k\times\nv)\star\vv_k^*]\cdot\vec{\la}\right\}. 
\end{equation}
We now want to show that the complex conjugate of the coefficient of 
$\vec{\la}$ is real by showing it is equal to itself 
($z^*=z \Rightarrow z$ is real).  Let us take the complex conjugate of 
the coefficient recalling that $\nv$ is real:
\begin{equation}
\frac{-ib}{c}\left\{[(\nv\times\vv_k)\star\vv_k^*]\cdot \vec{\la} 
          + [(\vv_k^*\times\vv_k)\star\nv]\cdot\vec{\la}
          + [(\vv_k^*\times\nv)\star\vv_k]\cdot\vec{\la}\right\}. 
\end{equation}
The first term here is the negative of the third term above, the second 
is the negative of the second above and the third is the negative of 
the first.  Therefore 
the two are equal due to the overall minus sign in this latter 
expression and so the coefficient is real and this part of the map is
real.  To prove that this sufficient condition is also necessary, we 
use the fact that there exists a minimal decompostion of the map 
such that each $T_k$ is independent and can always be put in minimal 
form \cite{Ou/Byrd:10a}.  Then, since the $\eta_k$ are real, this 
condition is also necessary when the map is in minimal form.  
$\square$


\subsubsection{Conditions for $T$ to be a symmetric matrix}

The importance of this stems from the fact that a symmetric matrix can
be decomposed in a polar decomposition 
\footnote{See also Section \ref{sec:svd} for further discussion.}
\begin{equation}
S = {\cal O} D {\cal O},
\end{equation}
where $S$ is a real symmetric matrix and ${\cal O}$ is orthogonal.  
We will show this to be true for a given $k$ and the
generalization follows from the sum $\sum_kT_k$.  

The only antisymmetric term in the expression for $T$ is
\begin{equation}
T^{(a)}_{pq} = \sum_k\eta_k if_{rpq}(v_{0k}^*v_{r k} - v_{0k}v_{r k}^*)
\end{equation}
Therefore, for $T$ to be symmetric, we require this term to vanish.  

Again, let us first consider the necessary condition that each term
given by a fixed $k$ vanish independently.  
Since the matrices $(f_{\alpha\beta})_\gamma$ form a
representation of the Lie algebra of SU(d), the only way for this to
to happen is if all vector components, indexed by $\gamma$, vanish
independently.  This implies that we must have 
\begin{equation}
v_{0k}^*v_{\gamma k} = v_{0k}v_{\gamma k}^*.
\end{equation}
There are several ways in which this can happen.  Let us consider some
examples.  Any of the following is sufficient to ensure that $T$ is symmetric:
\begin{enumerate}

\item $v_{0k}$ and $v_{\gamma k}$ are all real.  (Thus all $C_k$ are
  Hermitian.) 

\item  For a given $k$ either $v_{0k} = 0,$ or $v_{\gamma k} =0, \;
  \forall \gamma$. (Note $\vv_{\gamma k} =0$ for all $\gamma$ is
  rather trivial because then $C_k\propto \Bid$.)  

\item $v_{\gamma k}= v_{\delta k}, \; \forall k, \forall
  \gamma,\delta$ and $v_{0 k} =v_{\gamma k}$.  Note that the vectors
  can be different for each $k$, but all have the same components for
  each $k$.  They are also NOT required to be real.  

\end{enumerate}



\section{Example Channels}

\label{sec:examples}

Here we consider some examples of maps which are of special interest.  
These are far from exhaustive and we will consider other examples in 
future applications.  


\subsection{$C_k$ hermitian}

When the $C_k$ are hermitian, the $T$ matrix is symmetric as shown 
in the previous section.  Here we provide explicit expressions.  

Consider a map with $C_k = C_k^\dagger$ so that the $C_k$ are Hermitian, or
sometimes called self-adjoint.  Then $C_k$ can be expressed as 
\begin{equation}
C_k = (v_{0k}\Bid + v_{ik}\lambda_i),
\end{equation}
with $v_{0k}, v_{ik} \in \mathbb{R}$.  

Since the components of $T$, Eq.~(\ref{eq:Tcomponents}), are real, 
the matrix components of the affine map for Hermitian $C_k$, are
\bea
(T_k)_{pq} &=& |v_{0k}|^2 \delta_{pq}+\frac{2}{d}(\vv_k)_p(\vv_k )_q +2v_{0k}d_{rpq}(\vv_k )_r 
        \nonumber \\
     && -f_{tpr}f_{rsq}(\vv_k)_t(\vv_k )_s
              +d_{tpr}d_{rsq}(\vv_k)_t(\vv _k)_s,
\eea
where the third and fifth terms in Eq.~(\ref{eq:Tcomponents}) vanish since 
$v_{0k},\vv_k$ are real.  This is symmetric (in $p$ and $q$) since the 
first two terms clearly are symmetric and the latter two can be 
shown to be symmetric by renaming the dummy indices $s$ and $t$, and 
cyclicly permuting the indices on $c$ and $d$, viz.
\bea
f_{tpr}f_{rsq}(\vv_k)_t(\vv_k )_s &=& f_{spr}f_{rtq}(\vv_k)_s(\vv_k )_t
                          \nonumber \\
			         &=& f_{rsp}f_{tqr}(\vv_k)_s(\vv_k )_t
			  \nonumber \\
			          &=& f_{tqr} f_{rsp}(\vv_k)_t(\vv_k )_s.
\eea
Therefore if the $C_k$ are Hermitian, then the map $T_{pq}$ is 
real and symmetric in $p$ and $q$.

The affine part of the transformation reduces to
\begin{equation}
\vec{t}_k = \frac{1}{b}(2 v_{0k}\vv_k + \frac{1}{c} \vv_k\star \vv_k).
\end{equation}


\subsection{$C_k$ Unitary}

This case will be explored further in the next subsection where numerous 
reasons are given for the special consideration.  Here we 
provide an explicit expression for the $T$ matrix and note that $\vec{t}$ is 
zero for this case.  One case of interest is when the map is 
given by $\Phi(\rho) = \sum_k\eta_k U_k\rho U_k^\dagger$ where 
$\sum_k\eta_k =1$ and $\eta_k>0$.  

Consider a map with $C_k C_k^\dagger = \Bid = C_k^\dagger C_k$.  
Then $C_k$ can be expressed as 
\begin{equation}
C_k = (v_{0k}\Bid + v_{ik}\lambda_i),
\end{equation}
with the following constraints 
\begin{equation}
  \left(|v_{0k}|^2 + \frac{2}{d}\vec{v}_k\cdot\vec{v}_k^*\right)= 1,
\end{equation}
and 
\bea
0 &=& \left(v_{0k}\vec{v}_k^*+v_{0k}^*\vec{v}_k+i\vec{v}_k\times\vec{v}_k^* 
		  + \frac{1}{c}\vec{v}_k\star\vec{v}_k^*\right)
                \nonumber \\
  &=& \left(v_{0k}\vec{v}_k^*+v_{0k}^*\vec{v}_k+i\vec{v}_k^*\times\vec{v}_k 
		  + \frac{1}{c}\vec{v}_k\star\vec{v}_k^*\right).
\eea
This implies separately that 
\begin{equation}
0 = \left(v_{0k}\vec{v}_k^*+v_{0k}^*\vec{v}_k 
		  + \frac{1}{c}\vec{v}_k\star\vec{v}_k^*\right)
\end{equation}
and 
\begin{equation}
0 = \vec{v}_k\times\vec{v}_k^*.
\end{equation}

This implies that $T_k$ reduces to 
\bea
(T_k)_{pq} &=& |v_{0k}|^2 \delta_{pq}+\frac{2}{d}(\vv_k)_p(\vv_k^*)_q 
    +i[v_{0k}^*f_{rpq}(\vv_k)_r -v_{0k}f_{rpq}(\vv_k^*)_r]
         \nonumber \\
    && +[v_{0k}d_{rpq}(\vv_k^*)_r + v_{0k}^*d_{rpq}(\vv_k)_r]
        +i\big[d_{spr}f_{rtq}(\vv_k)_s(\vv_k^*)_t
                           +d_{rsq}f_{tpr}(\vv_k)_t(\vv_k^*)_s\big] 
         \nonumber \\
     && -f_{tpr}f_{rsq}(\vv_k)_t(\vv_k^*)_s
              +d_{tpr}d_{rsq}(\vv_k)_t(\vv^*_k)_s.
\label{eq:UnitaryTcomponents}
\eea


\subsection{Qutrits}

In this section we provide explicit example channels which are 
relevant for physical problems involving three-state systems.  
See also Ref.~\cite{Checinska/Wodkiewicz:08} and references therein.


\subsubsection{Qutrit Depolarizing Channel}

In general the effect of the depolarizing channel on a density
operator is to cause a uniform shrinking of the polarization vector
$\nv$.  This is written as
\beq
\rho = \frac{1}{d}(\Bid + b \nv\cdot\lav) \rightarrow 
\rho^\prime =  \frac{1}{d}(\Bid + b p\nv\cdot\lav),
\eeq
where $p$ is the shrinking factor.  (See for example
\cite{Byrd/Brennen:08} and references therein.) 

The effect of the depolarizing channel 
on the coherence vector is such that it uniformly shrinks each component 
of $\nv$ by a common factor. It can be readily verified that if we let 
\begin{equation}
C_0 = \sqrt{1-x}\:\Bid,
\end{equation}
and
\begin{equation}
C_k = \sqrt{3x/16} \: \lambda_k, \;\;\;k=1,2,\hdots,8
\end{equation}
we then have $\sum_{k=0}^8 C_k C_k^{\dagger} = \Bid$, 
$\vec{t}=0$, $(T_0)_{pq}=(1-x)\delta_{pq}$, and 
$(T_k)_{pq}= (x/8)\delta_{kp}\delta_{kq} +(3x/16)(d_{kpr}d_{rkq} - 
f_{kpr}f_{rkq})$ for 
$k=1,2,\hdots,8.$ The coherence vector is thus transformed 
according to 
\begin{equation}
\nv \mapsto  T\nv = \sum_k T_k \nv = (1 - 9x/8) \nv.
\end{equation} 
Thus the shrinking factor is $p= 1-9x/8$.


\subsubsection{Qutrit Phase Damping Channel}

Analogous to the qubit phase damping channel, the effect of the 
phase damping channel on a three state system is such that it 
leaves the diagonal components of $\nv$ unchanged while uniformily 
shrinking the off-diagonal components. When the operators 
$C_k$ are chosen to be 
\begin{equation}
C_0 = \sqrt{1-x}\:\Bid,
\end{equation}
\begin{equation}
C_1 = \sqrt{3x/20}\:(\Bid + \lambda_3),
\;\;\;
C_2 = \sqrt{3x/20}\:(\Bid - \lambda_3),
\end{equation}
and
\begin{equation}
C_3 = \sqrt{3x/20}\:(\Bid + \lambda_8),
\;\;\;
C_4 = \sqrt{3x/20}\:(\Bid - \lambda_8),
\end{equation}
it can be shown that $\sum_{k=0}^4 C_k C_k^{\dagger} = \Bid$, 
$\vec{t}=0$, $(T_0)_{pq}=(1-x)\delta_{pq}$, $(T_1)_{pq} = \mu_3^+$, 
$(T_2)_{pq} = \mu_3^-$, $(T_3)_{pq} = \mu_8^+$, and $(T_4)_{pq} = \mu_8^-$,
where
\begin{equation}
\mu_l^{\pm} \equiv \frac{3x}{20} \left[\delta_{pq} + \frac{2}{3}\delta_{pl}
\delta_{ql} \pm d_{lpq} - f_{lpr}f_{lqr} + d_{lpr}d_{lqr} \right].
\end{equation}
When $\eta_k = 1$ for $\forall k$ the matrix T shrinks the off-diagonal 
components of the coherence vector by a factor of $1-3x/5$, i.e.,
\begin{equation}
n_i 
\mapsto \left\{\begin{array}{ll}  
n_i,  & 
\mbox{for} \;\; i = 3,8 \\
(1-3x/5)n_i,  &
\mbox{otherwise}  
\end{array}\right.
\end{equation}
In the last expression we have taken the usual convention of labeling 
the diagonal elements of the basis as 3 and 8.


\subsubsection{Off-Diagonal Channel}

Let us now define 
\begin{equation}
C_0 = \sqrt{1-x}\:\Bid,
\end{equation}
and
\begin{equation}
C_k = \sqrt{x/4}\: \lambda_k, \;\;\; \mbox{for} \;\;\; k=1,2,4,5,6,7.
\end{equation}
These values of $k$ correspond to the off-diagonal generators of 
$SU(3)$. It can be verified that this particular OSR transforms 
the coherence vector according to
\begin{equation}
n_i 
\mapsto \left\{\begin{array}{ll}  
(1-3x/2)n_i,  & 
\mbox{for} \;\; i = 3,8 \\
(1-x)n_i,  &
\mbox{otherwise}  
\end{array}\right.
\end{equation}
when $\eta_k = 1$ for $\forall k$. In this case, the diagonal components 
of $\nv$ shrink faster with $x$ than the off-diagonal components. 
If instead $\eta_k = (-1)^k$, the mapping given by 
$T=\sum_k (-1)^k T_k \:\:\:($again $\tv = \sum_k (-1)^k \tv_k = \vec{0})$ 
affects the components of $\nv$ in such a way as to shrink them according 
to   
\begin{equation}
n_i 
\mapsto \left\{\begin{array}{ll}  
(1-3x/2)n_i,  & 
\mbox{for} \;\; i = 1,4,6 \\
(1-x/2)n_i,  &
\mbox{for} \;\; i = 2,5,7 \\
(1-x)n_i,  &
\mbox{for} \;\; i = 3,8 
\end{array}\right.
\end{equation}

The components of $\nv$ associated with the real off-diagonal 
elements $(\lambda_i, \; i = 1,4,6)$ shrink with 
$x$ faster than those associated with the imaginary off-diagonal elements 
$(\lambda_i, \; i = 2,5,7)$. The diagonal components are reduced in 
magnitude at an intermediate rate. The values assigned to 
the $\eta_k$ have a significant effect on the 
evolved state, as clearly illustrated by this example. 


\subsubsection{Trit Flip Channel}

For qubits, the bit flip channel acts in such a way as to flip the basis state 
$\ket{1}$ to $\ket{2}$ and vice versa. For qutrits there are three basis 
states which can be labeled as $\ket{1}$, $\ket{2}$, and $\ket{3}$. To 
describe the effect of flipping one of these basis states to another in 
terms of an affine mapping of the polarization vector let us first 
define the following pure state density matrices
\beq 
\rho_1 =
\left(\begin{array}{ccc}
      1 & 0 & 0  \\
      0 & 0 & 0  \\
      0 & 0 & 0  \\
      
        \end{array}\right), \;\;\;
\rho_2 =
\left(\begin{array}{ccc}
      0 & 0 & 0  \\
      0 & 1 & 0  \\
      0 & 0 & 0  \\
       \end{array}\right),   \;\;\;
\rho_3 =
\left(\begin{array}{ccc}
      0 & 0 & 0  \\
      0 & 0 & 0  \\
      0 & 0 & 1  \\
       \end{array}\right).  
\eeq
Now the affect of flipping from the state $\ket{1}$ to $\ket{2}$ is 
equivalent to the density operator $\rho_1$ evolving to $\rho_2$. 
This can be acheived by the transformation 
$C_{(1,2)}^{\phantom{\dagger}}\rho_1 C_{(1,2)}^{\dagger} = \rho_2$,
where
\beq 
C_{(1,2)} =
\left(\begin{array}{ccc}
      0 & 1 & 0  \\
      1 & 0 & 0  \\
      0 & 0 & 0  \\
      \end{array}\right).
\eeq
Notice that $C_{(1,2)}$ is an element of the Gell-Mann basis for $SU(3)$. 
The affine map associated with this transformation has a linear part as well 
as a nonzero translational part, specifically
\beq 
T_{(1,2)} =
\left(\begin{array}{cccccccc}
      1 & 0 & 0 & 0 & 0 & 0 & 0 & 0\\
      0 & -1 & 0 & 0 & 0 & 0 & 0 & 0  \\
      0 & 0 & -1 & 0 & 0 & 0 & 0 & 0 \\
      0 & 0 & 0 & 0 & 0 & 0 & 0 & 0 \\
      0 & 0 & 0 & 0 & 0 & 0 & 0 & 0 \\
      0 & 0 & 0 & 0 & 0 & 0 & 0 & 0 \\
      0 & 0 & 0 & 0 & 0 & 0 & 0 & 0 \\
      0 & 0 & 0 & 0 & 0 & 0 & 0 & 1/3 \\
      \end{array}\right), \;\;\; {\mbox{and}} \;\;\;
t_{(1,2)} =
\left(\begin{array}{c}
      0 \\
      0 \\
      0 \\
      0 \\
      0 \\
      0 \\
      0 \\
      1/3 \\
      \end{array}\right).
\eeq
Note that since $C_{(1,2)}^{\phantom{\dagger}}\rho_2 C_{(1,2)}^{\dagger} 
= \rho_1$, the affine map associated with the flip  
$\ket{2} \rightarrow \ket{1}$ is the same as above. Similarly one can
show that the transformation 
$C_{(1,3)}^{\phantom{\dagger}}\rho_1 C_{(1,3)}^{\dagger} = \rho_3$, with
\beq 
C_{(1,3)} =
\left(\begin{array}{ccc}
      0 & 0 & 1  \\
      0 & 0 & 0  \\
      1 & 0 & 0  \\
      \end{array}\right),
\eeq
is equivalent to the affine mapping
\beq 
T_{(1,3)} =
\left(\begin{array}{cccccccc}
      0 & 0 & 0 & 0 & 0 & 0 & 0 & 0\\
      0 & 0 & 0 & 0 & 0 & 0 & 0 & 0  \\
      0 & 0 & 0 & 0 & 0 & 0 & 0 & -1/\sqrt{3} \\
      0 & 0 & 0 & 1 & 0 & 0 & 0 & 0 \\
      0 & 0 & 0 & 0 & -1 & 0 & 0 & 0 \\
      0 & 0 & 0 & 0 & 0 & 0 & 0 & 0 \\
      0 & 0 & 0 & 0 & 0 & 0 & 0 & 0 \\
      0 & 0 & -1/\sqrt{3} & 0 & 0 & 0 & 0 & -2/3 \\
      \end{array}\right), \;\;\; {\mbox{and}} \;\;\;
t_{(1,3)} =
\left(\begin{array}{c}
      0 \\
      0 \\
      1/2\sqrt{3} \\
      0 \\
      0 \\
      0 \\
      0 \\
      -1/6 \\
      \end{array}\right). 
\eeq
Since $C_{(1,3)}^{\phantom{\dagger}}\rho_3 C_{(1,3)}^{\dagger} = \rho_1$,
the mapping associated with $T_{(1,3)}$ and $t_{(1,3)}$ will yield the 
transition $\ket{3} 
\rightarrow \ket{1}$. For the flips $\ket{2} 
\leftrightarrow \ket{3}$, the associated OSR decomposition has 
the term $C_{(2,3)}$
with
\beq 
C_{(2,3)} =
\left(\begin{array}{ccc}
      0 & 0 & 0  \\
      0 & 0 & 1  \\
      0 & 1 & 0  \\
      \end{array}\right).
\eeq
This leads to an affine mapping of the form
\beq 
T_{(2,3)} =
\left(\begin{array}{cccccccc}
      0 & 0 & 0 & 0 & 0 & 0 & 0 & 0\\
      0 & 0 & 0 & 0 & 0 & 0 & 0 & 0  \\
      0 & 0 & 0 & 0 & 0 & 0 & 0 & 1/\sqrt{3} \\
      0 & 0 & 0 & 0 & 0 & 0 & 0 & 0 \\
      0 & 0 & 0 & 0 & 0 & 0 & 0 & 0 \\
      0 & 0 & 0 & 0 & 0 & 1 & 0 & 0 \\
      0 & 0 & 0 & 0 & 0 & 0 & -1 & 0 \\
      0 & 0 & 1/\sqrt{3} & 0 & 0 & 0 & 0 & -2/3 \\
      \end{array}\right), \;\;\; {\mbox{and}} \;\;\;
t_{(2,3)} =
\left(\begin{array}{c}
      0 \\
      0 \\
      -1/2\sqrt{3} \\
      0 \\
      0 \\
      0 \\
      0 \\
      -1/6 \\
      \end{array}\right),
\eeq


\section{A preferred basis: $C_k$ unitary and orthogonal}
\label{sec:unandortho}

It is often the case that a preferred basis for the $C_k$ is chosen.
One important case, is when the $C_k$ are both 
orthogonal and unitary.  This choice corresponds
to a nice error basis \cite{Knill/LANL:96} and applies to
the so-called generalized Pauli matrices which are formed from 
$C_k$ given by $U_{m,n}=X^mZ^n$,
$m, n=0, 1, 2$ where 
$X|j\rangle=|j+1 \texttt{mod} d\rangle$,
$Z|j\rangle=w^j|j\rangle$, $w=e^{2\pi i/3}$ and 
$\{|j\rangle\}_{j=0}^2$ is an orthonormal basis. We also note that it
can be applied to decoupling controls which form an discrete subgroup
of the unitary group, but in that case the Hamiltonian is modified
rather than the density operator \cite{Ou/Byrd:tpb}.

Since the case where the $C_k$ are unitary and orthogonal is a
particularly important case, we show that when the $C_k$ are
orthogonal, unitary, and belong to a 
discrete subgroup of the unitary group, the trace of pairs of unequal
$T_k$ must be $-1$ between elements of the set.  

Let $\{U_k\}$ be such a set which also forms a subset of 
a fundamental irreducible
representation of a discrete subgroup of the unitary group, i.e.,
$U_k^\dagger U_k =\Bid$, and the set of $U_k$ can be extended to form 
a discrete subgroup of the unitary group, and 
Tr$(U_iU_j^\dagger)=0$ for $i\neq j$.  Then, since $U_j^\dagger U_i$
is a group element, the trace corresponds to the character of the
product.  The character, for an irreducible representation, is the
same for each element in a class and, in this case, the class has 
character equal to zero.  However, it can be different for different
representations.  Note that the corresponding $T$ matrix is
an orthogonal matrix since it belongs to the adjoint representation of
the group which is a real $(d^2-1)$-dimensional representation.  This
follows from the equation
\beq
U_k\rho U_k^\dagger = U_k\frac{1}{d}\left(\Bid + b\nv\cdot\lav\right)U_k^\dagger 
                = \frac{1}{d}\left(\Bid + b\sum_{ij}n_i(R_k)_{ij}\lambda_j\right),
\eeq
where $U_k\in U(d)$ and $R_k$ is in the adjoint representation of the
group.  In this case $T_k$ corresponds to $R_k$, and $U_k$ is the same
group element as $R_k$ but in a different representation.  

Since $\tr(U^\dagger_iU_j)=0$, and $U_i\otimes U_j$ is in the tensor
product of the representation of $U(d)\otimes U(d)$, we can write
$\tr(U_i^\dagger\otimes U_j^\dagger U_k\otimes U_l) = 0$ and this is
the product of the two representations.  If we take the complex
conjugate of the first of these two representations (to obtain the
conjugate representation) we find that the
product, which is equal to the direct sum of the real 
$(d^2-1)$-dimensional representation and the trivial one, must be
zero.  Since the trace of the sum is the sum of the traces, and the
trivial representation has character (trace) one, the other must have
trace $-1$.  Therefore, if the $C_k$ are unitary, orthogonal, and
the set of these can be extended to form a group, 
then the corresponding $T_k$ have the traces of the 
products all equal to $-1$, viz., 
$\tr(C_iC_j^\dagger) =0 \Leftrightarrow \tr(T_i^TT_j)=-1, \; \forall \;
i\neq j$.  

As mentioned above, this result is important for error prevention and
control methods.  It is also generally important for translating
between the OSR and the affine map when this basis is chosen.  

We now briefly discuss how one would obtain the affine map directly
from the dynamical map, showing that the OSR is in fact not a
necessary intermediate step.  


\section{Affine Map from the Dynamical Map}
\label{sec:dyntoOSR}

It is fairly straight-forward to show that the affine map can be
obtained directly from the dynamical map.  In this case, a convenient
basis is the set of matrices which form a 
basis for GL($n,\mathbb{C}$).  Such a basis is 
defined by the set of matrices with a $1$ in the $i$th column and
$j$th row and zeros everywhere else, denoted $\{E_{ij}\}$.  Given that
from 
Eq.~(\ref{eq:As1}),
we see that the two bases may be converted from one to another using, 
for example, the correspodence between the raising and lower operators
of the group and the hermitian basis.  To be specific, one would expand
$\rho = \sum_{r,s}\rho_{rs} E_{rs}$, and use the corresponding
hermitian basis $\Lambda_{rs} = E_{rs}+E_{sr}$ and
$\Lambda_{rs}^\prime = i(E_{rs}-E_{sr})$ as well as linear
combinations of the identity and Cartan algebra elements to expand the
diagonal matrices.  (See for example \cite{Bishop/Byrd:09} and
references therein.)  

Thus we have given the following correspondences: dynamical maps
$\rightarrow$ OSR $\rightarrow$ affine map and dynamical map
$\rightarrow$ affine map.  Note, however, that the OSR is not unique
-- not for completely positive matrices \cite{Nielsen/Chuang:book} nor for
those that are not completely positive \cite{Ou/Byrd:10a}.  Therefore,
the set of $T_k$ matrices is also not unique and thus converting from
one picture to another is not a one-to-one transformation.  The
dynamical map is essentially unique as is the $T$ matrix as well as
the minimal decomposition of the map \footnote{This lack of uniqueness
  is explained in \cite{Ou/Byrd:10a}.  See also references therein.}.


\section{Singular Value Decomposition}

\label{sec:svd}

In the analysis of maps of qubit density operators, sufficient 
conditions for the positivity of maps follows from 
positivity conditions described by the Bloch-sphere
\cite{King/Ruskai,Fujiwara/Algoet:99}.  
The maps, represented as an affine map of
the vector, can be decomposed in
terms of a singular value decomposition (SVD) which greatly
simplifies the analysis 
\cite{King/Ruskai}.  Here we discuss a decomposition which is similar
to the singular value decomposition of 
the affine map for qubits, but applies to a $d$-dimensional system
when $d>2$.  This turns out to be much more complicated for $d$-state
systems than it is for a two-state system.  The reason, as explained
in detail below, is that there are restrictions on the ``rotation'' part of 
the map (that part which does not change the magnitude of $\nv$).  
The restrictions are determined by the ``shrinking'' part of the map
(that part which can reduce the magnitude of one or more components of
$\nv$) and differ for different shrinking matrices.  
Thus we are seeking an answer to the following question.  To what
extent can we generalize the results of qubits to qudits with $d>2$?  
Our results provide conditions under which a SVD may be performed as well as 
restrictions to its use.  This provides some insight into the
geometry of the space.


\subsection{SVD For Affine Qubit Maps}

Before presenting our results for qudits, we will review the basic
idea for qubits as used in Ref.~\cite{King/Ruskai}.  Let some 
initial density matrix 
\begin{equation}
\rho = \frac{1}{2}(\Bid + \nv\cdot \sigmav),
\end{equation}
be acted on by  
\begin{equation}
\Phi(\rho) = \frac{1}{2}(\Bid + \nv^\prime \cdot \sigmav).
\end{equation}
This map can be described by an affine map of the polarization vector,
\begin{equation}
\nv^\prime = T\nv + \tv.
\end{equation}
In what follows we will let $\tv=0$ and discuss $T$.  The translation
may be treated separately.

For two-state systems, maps can be diagonalized by means of a SVD.
This is because any real $N\times N$ matrix $M$ can be written as
\begin{equation}
M = {\cal O}_1 D {\cal O}_2,
\end{equation}
where ${\cal O}_i\in SO(N)$ and $D$ is a diagonal matrix with the
entries being called the singular values.  For qubit maps $N=d^2-1$ is
3 and an action of $U\in U(2)$ on the density operator is equivalent
to an $SO(3)$ matrix acting on the polarization vector $\nv$.  To see
this, we write 
\beq
U\rho U^\dagger = \frac{1}{2}\left(\Bid +
  \sum_{ij}n_iR_{ij}\sigma_j\right), 
\eeq
where $R\in SO(3)$.  Thus acting before and after a map enables one to
choose a preferred basis for the map.  This basis can be chosen so
that $T$ is diagonal since choosing $R$ and $R^\prime$ appropriately
enables us to put $T$ into a diagonal form using $D = RTR^\prime$.   
This set of diagonalized maps can be considered a double coset space
of the elements of the set of maps.  

For future reference it is relevant to note that for a positive map,
the matrix $D$ cannot increase the magnitude of $\nv$ or any component
of $\nv$.  (Again, we are taking $\tv =0$.)  
In general, for the SVDs we will consider here, we will refer to the
$D$ part (diagonal part) as the ``shrinking factor'' since, for a
positive map, it can decrease, but not increase, the magnitude of the
polarization vector.  The parts ${\cal O}$ will be
called ``rotations'' although the maps can clearly be physically
achieved in many different ways.


\subsection{SVD For Qutrits}

Here we will discuss the case of qutrits.  The generalization to qudits
follows.  We know that for a
mapping from pure states to pure states, the map is a unitary
transformation.  That is, any map from a pure state to a pure state
can be written as a unitary transformation.  The unitary
transformation acts on the density operator to produce an affine map
which has the form $T\in$Ad(SU(3))$\subset$SO(8).  That is, the $T$
matrix is an element of the adjoint representation of SU(3) which is
a subset of SO(8).  It clearly does not span the space of SO(8) in
the way that SU(2) covers SO(3) since the dimension of SU(3) is
$3^2-1$ = 8 and the dimension of SO(8) is $8\cdot 7/2=28$.

For an even simpler case, consider a map which
acts as a depolarizing channel together with a ``rotation.''    
In this case there is a uniform
shrinking factor such that each component of the polarization vector
is reduced by the same factor $p<1$.  (See for example
\cite{Byrd/Brennen:08} and references therein.)  For qutrits, when
$p<1/2$, the map has a SVD with 
\begin{equation}
T = {\cal O}_1 D {\cal O}_2,
\end{equation}
where ${\cal O}_i\in$ SO(8).  Any matrix in SO(8) is allowed since
any direction for the polarization vector leads to a positive density
matrix output as long as the shrinking factor is $p<1/2$ and uniform.  
This follows from the positivity conditions in
Refs.~\cite{Byrd/Khaneja:03,Kimura}.  

For there to exist a SVD for maps which have $1/2<p<1$
the SVD can not either be the SVD for
$p<1/2$ nor can it be the SVD for pure states.  The SVD for these
intermediate values of $p$ must have some elements ${\cal O}_i\in$
SO(8) that can not be arbitrary elements of SO(8), but come from a
restricted set which depend on the shrinking factor.  
However, they are also not restriced only to the set
of matrices in SU(3)$\subset$ SO(8), they are from a larger set.  The
question is, what is that set and how do we find the values?  

Whereas the general problem is quite complicated, we can make some
general statements about the allowable rotations.  Let us consider
the algebra of SU(3)$\subset$ SO(8).  For pure states, the set of
allowable rotations are only those which are in SU(3) and at the other
extreme, all rotations in SO(8) are allowable when the magnitude of
the polarization vector $|\nv|\leq 1/2$ as discussed above.  
Thus the SO(8) symmetry is
much larger than the SU(3) symmetry and it could be said that the
symmetry is broken continuously by the increasing magnitude of the
polarization vector.  Such considerations are important for a
    variety of reasons.  Most notably, the ``good quantum numbers''
    are labels for states which are carriers of an irreducible
    representation of a group.  (See for example \cite{Bohm:qmbook}.)  
    Thus symmetries determine 
    simultaneously measurable quantities for a quantum system and thus
  enable us to specify a state as well as we can.  When a symmetry is
  broken, the symmetry for the system is reduced and the set of
  good quantum numbers changes, indicating a change in allowed states
  of the system.  In this description of the allowable states,  
one possible symmetry breaking construction
which could be used for broken symmetries is the
following subgroup chain:
$$
SO(8) \supset SO(6)\times SO(2) \cong SU(4)/\mathbb{Z}_2 \times U(1)
\supset SU(3).  
$$
Note, however, that the restricted symmetry arises from positivity
conditions on the density operator.  This positivity condition is
associated with purity/mixedness of the system.  Thus we have a
connection between the positivity of the system and its entropy.  This
relation is only direct for two-state systems, however, since there are
two positivity constraints for three-level systems and more for
higher-dimensional systems.  

The explicit parameterization of the rotation matrices can be obtained
from the exponentiation of the algebraic elements given in the
appendix.  There we have expressed the algebraic elements of SO(8) in
terms of the algebraic elements of SU(3).  This embedding of the SU(3)
rotations into the set of SO(8) rotations is important since it
specifies the unitary part of the transformation.  We will not
provide the explicit parameterizations in each of the regimes, but do
discuss an example.  

The general case is straight-forward to infer.  There is a ball at the
center of the space of density operators which has a spherical
symmetry in which all density operators are positive semi-definite.
At the other extreme are the pure states which can only be acted upon
by a rotation part consisting of unitary transformations.  The
intermediate points have ``rotation'' matrices which are restricted
and depend on the magnitude and direction of the polarization vector.  

As an illustrative example we consider the case where a 
qutrit is initially in the pure state $\ket{2}$ (let $\ket{1}$,$\ket{2}$, 
and $\ket{3}$ represent the basis states of a three-state system). The 
coherence vector associated with the density 
matrix $\ket{2} \!\!\bra{2}$ is given by 
$\nv = (0,0,-\sqrt{3}/2,0,0,0,0,1/2)^T$, where the 
third and eighth components are associated with the 
diagonal generators of $SU(3)$, i.e., $\ket{1} \!\!\bra{1} 
- \ket{2} \!\!\bra{2}$ and $1/\sqrt{3}(\ket{1} \!\!\bra{1} + 
\ket{2} \!\!\bra{2} -2\ket{3} \!\!\bra{3})$ respectively. If the components 
of $\nv$ experience a rotation due to, say, the orthogonal matrix $f_{28}$ 
(see Appendix B), and shrink in magnitude by the factor $(1-p)$, the evolved 
state $\nv^{\prime} = (1-p)\exp{(-i\theta f_{28})}\nv$ may or may not 
correspond to a positive-semidefinite matrix. Since the initial 
state was pure, the necessary condition for positivity is 
$S_3 = 1-3\nv^{\prime}\cdot\nv^{\prime}+2 (\nv^{\prime} \star \nv^{\prime})
\cdot \nv^{\prime} \geq 0$, see Ref.~\cite{Byrd/Khaneja:03,Kimura}.

\begin{figure}[!ht]
\includegraphics[width=.4\textwidth]
{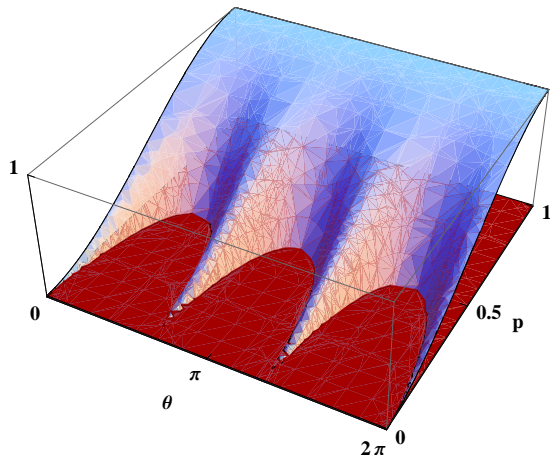}
\caption{Plot showing the region of positivity when the initial pure 
state $\ket{2}$ is allowed to experience a uniform shrinking $(1-p)$ and 
rotation of angle $\theta$ by the matrix $f_{28}$ (see Appendix B). The quantity $1-3\nv^{\prime}\cdot\nv^{\prime}+2 (\nv^{\prime} \star \nv^{\prime})
\cdot \nv^{\prime}$ is graphed vertically since for this initial pure state 
the condition $1-3\nv^{\prime}\cdot\nv^{\prime}+2 (\nv^{\prime} \star 
\nv^{\prime})
\cdot \nv^{\prime} \geq 0$ is necessary and sufficient for positivity.      }
\label{fig:pos}
\end{figure}

In Fig.~\ref{fig:pos} we plot the region of positivity as a function 
of the parameters $\theta$ and $p$. The figure shows that while 
$S_3 = 0$ initially $(p=\theta=0)$, this quantity quickly becomes negative as 
the state begins to rotate without shrinking. The figure also shows, although 
not very clearly, how the initial state becomes positive again as $\theta$ 
increases to multiples of $2\pi/3$ while no shrinking has occured. When 
the components of the initial coherence vector uniformily shrink by an amount 
$(1-p) > 1/2$ we see that any rotation about the matrix $f_{28}$ leads to 
a positive density operator.


\section{Concluding Remarks}
\label{sec:concl}

We have obtained an expression for an arbitrary affine mapping of the 
polarization vector associated with a $d$-dimensional density operator 
in terms of the components of the operator-sum 
representation of the dynamical map.  This provides a geometric
picture, albeit somewhat abstract, via the polarization vector
components which are measurable quantities.  We also described a
direct method for expressing the affine map in terms of the dynamical
map.  Some example channels were provided in order to 
highlight the connection between particular terms in our expression and 
specific operators appearing in the OSR.  For the important case that
the OSR components are unitary and orthogonal, we have shown that the
corresponding affine components are real and have trace -1 between
pairs of unequal terms.  This is particularly useful for the
generalization of the Bloch equations since this basis is used for
error prevention methods.  

To generalize the methods used for qubit maps, we discussed  
the possible generalization of the singular value decomposition 
of qubit affine maps to higher dimensional systems.  We have found
that a symmetry-breaking occurs due to positivity constraints.  Thus
we have given a relation between 
the physical symmetry, the purity and the entropy of
the physical system.  In this particular case we have provided the
example of the depolarizing channel, but the generalization is also
discussed.  

While we have provided some particular expressions for physically
motivated quantum channels (maps), applications of this work range
from information theory to practical experiments as is the case with
the OSR. One example of the utility is given in Ref.~\cite{Wu/Byrd:algs},
where the robust simulation of quantum systems with quantum
systems was considered.  In addition, the preferred basis given here
provides a connection between quantum error correcting codes and
direct application thereof.  This is due, in part, to the ability to
more directly determine the input and output, and therefore the map
itself, for quantum systems undergoing noisy evolution.  In future
work we will use this to try to provide more practical methods for
determining the best combination of error prevention methods.  (See
\cite{Byrd/etal:pqe04} for a review.)

More generally, we anticipate applications of this could be more wide
spread in quantum control.  For example, classical control 
systems utilizing affine maps may be found 
in Judjevic's book, ``Geometric Control Theory,'' chapter 4.  This
includes Linear systems reachability, and constraints.  We also 
hope that our work will be useful in ways not
yet foreseen by us.


\appendix

\section{Identities}



\label{app:sunalg}

In this appendix we provide identities that were used in the
derivations of the expressions above.  

The commutation relations, anti-commutation relations, and
normalization of the matrices 
representing the basis for the Lie algebra can be summarized 
by the following equation:
\begin{equation}
\label{eq:larels}
\lambda_i \lambda_j = \frac{2}{d}\delta_{ij} + if _{ijk} \lambda_k 
                      + d_{ijk}\lambda_k,
\end{equation}
where here, and throughout this appendix, a sum over repeated 
indices is understood.  

As with any Lie algebra we have the Jacobi identity:
\begin{equation}
\label{eq:jid}
f_{ilm}f_{jkl} + f_{jlm}f_{kil} + f_{klm}f_{ijl} =0.
\end{equation}
There is also a Jacobi-like identity,
\begin{equation}
\label{eq:jlid}
f_{ilm}d_{jkl} + f_{jlm}d_{kil} + f_{klm}d_{ijl} =0,
\end{equation}
which was given by Macfarlane, et al. \cite{Macfarlane}. 

The following identities, are also provided in \cite{Macfarlane}, 
\begin{eqnarray}
d_{iik} &=& 0, \label{eq:Mac(2.7)}\\
d_{ijk}f_{ljk} &=& 0,  \label{eq:Mac(2.14)}\\
f_{ijk}f_{ljk} &=& d\delta_{il},  \label{eq:Mac(2.12)}\\
d_{ijk}d_{ljk} &=& \frac{d^2 - 4}{d}\delta_{il},  \label{eq:Mac(2.13)}
\end{eqnarray}
and
\begin{equation}
f_{ijm}f_{klm} = \frac{2}{d}(\delta_{ik}\delta_{jl} - \delta_{il}\delta_{jk}) 
                  + (d_{ikm}d_{jlm} - d_{jkm}d_{ilm}) \label{eq:Mac(2.10)}
\end{equation}
 and finally
\begin{eqnarray}
f_{piq}f_{qjr}f_{rkp} &=& -\left(\frac{d}{2}\right)f_{ijk}, \label{eq:Mac(2.15)}\\
d_{piq}f_{qjr}f_{rkp} &=& -\left(\frac{d}{2}\right)d_{ijk}, \label{eq:Mac(2.16)}\\
d_{piq}d_{qjr}f_{rkp} &=& \left(\frac{d^2 - 4}{2d}\right)f_{ijk}, \label{eq:Mac(2.17)}\\
d_{piq}d_{qjr}d_{rkp} &=& \left(\frac{d^2 - 12}{2d}\right)d_{ijk} \label{eq:Mac(2.18)}.
\end{eqnarray} 
The proofs of these are fairly straight-forward and are omitted.  


\section{Algebra of rotation matrices}

\label{app:rotation}

In this section we provide the basis elements for the symmetry
breaking from the SO(8) symmetry to the SU(3) symmetry.  A basis for
the algebra of SO(8) is given by one set of matrices and the SU(3)
algebraic basis elements are expressed in 
terms of that basis using the structure constants which provide the
algebra for the adjoint representation of the group.  This provides
the afforementioned embedding of the group SU(3) into SO(8) by
the exponentiation of the algebra.  

The structure constants for SU(3) are $f_{ijk}$ and we write them in
matrix form as $(f_i)_{jk}$, i.e. the matrix $f_i$ has elements $j,k$.  
Consider the basis for a matrix representation of SO(8) given by the
set of all 
$$
m_{jk} = i(\dmat{j}{k} - \dmat{k}{j}),
$$
for which $j<k=2,...,8$.  

For SU(3) we label the matrices 1-8, with elements $j,k$ corresponding
to the eight Gell-Mann matrices.  The basis is 
$$
\begin{array}{cc}
f_{1jk}=i\left(\begin{array}{cccccccc}
 0 & 0 & 0 & 0 & 0 & 0 & 0 & 0 \\
 0 & 0 & 1 & 0 & 0 & 0 & 0 & 0 \\
 0 & -1 & 0 & 0 & 0 & 0 & 0 & 0 \\
 0 & 0 & 0 & 0 & 0 & 0 & \half & 0 \\
 0 & 0 & 0 & 0 & 0 & -\half & 0 & 0 \\
 0 & 0 & 0 & 0 & \half & 0 & 0 & 0 \\
 0 & 0 & 0 & -\half & 0 & 0 & 0 & 0 \\
 0 & 0 & 0 & 0 & 0 & 0 & 0 & 0 \end{array}\right) & 
f_{2jk}=i\left(\begin{array}{cccccccc}
 0 & 0 & 1 & 0 & 0 & 0 & 0 & 0 \\
 0 & 0 & 0 & 0 & 0 & 0 & 0 & 0 \\
 -1 & 0 & 0 & 0 & 0 & 0 & 0 & 0 \\
 0 & 0 & 0 & 0 & 0 & \half & 0 & 0 \\
 0 & 0 & 0 & 0 & 0 & 0 & \half & 0 \\
 0 & 0 & 0 & -\half & 0 & 0 & 0 & 0 \\
 0 & 0 & 0 & 0 & -\half & 0 & 0 & 0 \\
 0 & 0 & 0 & 0 & 0 & 0 & 0 & 0 \end{array}\right) \\
f_{3jk}=i\left(\begin{array}{cccccccc}
 0 & 1 & 0 & 0 & 0 & 0 & 0 & 0 \\
 -1 & 0 & 0 & 0 & 0 & 0 & 0 & 0 \\
 0 & 0 & 0 & 0 & 0 & 0 & 0 & 0 \\
 0 & 0 & 0 & 0 & \half & 0 & 0 & 0 \\
 0 & 0 & 0 & -\half & 0 & 0 & 0 & 0 \\
 0 & 0 & 0 & 0 & 0 & 0 & -\half & 0 \\
 0 & 0 & 0 & 0 & 0 & \half & 0 & 0 \\
 0 & 0 & 0 & 0 & 0 & 0 & 0 & 0 \end{array}\right) &
f_{4jk}=i\left(\begin{array}{cccccccc}
 0 & 0 & 0 & 0 & 0 & 0 & -\half & 0 \\
 0 & 0 & 0 & 0 & 0 & -\half & 0 & 0 \\
 0 & 0 & 0 & 0 & -\half & 0 & 0 & 0 \\
 0 & 0 & 0 & 0 & 0 & 0 & 0 & 0 \\
 0 & 0 & \half & 0 & 0 & 0 & 0 & \frac{\sqrt{3}}{2} \\
 0 & \half & 0 & 0 & 0 & 0 & 0 & 0 \\
 \half & 0 & 0 & 0 & 0 & 0 & 0 & 0 \\
 0 & 0 & 0 & 0 & -\frac{\sqrt{3}}{2} & 0 & 0 & 0 \end{array}\right) \\
f_{5jk}=i\left(\begin{array}{cccccccc}
 0 & 0 & 0 & 0 & 0 & \half & 0 & 0 \\
 0 & 0 & 0 & 0 & 0 & 0 & -\half & 0 \\
 0 & 0 & 0 & \half & 0 & 0 & 0 & 0 \\
 0 & 0 & -\half & 0 & 0 & 0 & 0 & \frac{\sqrt{3}}{2} \\
 0 & 0 & 0 & 0 & 0 & 0 & 0 & 0 \\
 -\half & 0 & 0 & 0 & 0 & 0 & 0 & 0 \\
 0 & \half & 0 & 0 & 0 & 0 & 0 & 0 \\
 0 & 0 & 0 & \frac{\sqrt{3}}{2} & 0 & 0 & 0 & 0 \end{array}\right) &
f_{6jk}=i\left(\begin{array}{cccccccc}
 0 & 0 & 0 & 0 & -\half & 0 & 0 & 0 \\
 0 & 0 & 0 & -\half & 0 & 0 & 0 & 0 \\
 0 & 0 & 0 & 0 & 0 & 0 & \half & 0 \\
 0 & \half & 0 & 0 & 0 & 0 & 0 & 0 \\
 \half & 0 & 0 & 0 & 0 & 0 & 0 & 0 \\
 0 & 0 & 0 & 0 & 0 & 0 & 0 & 0 \\
 0 & 0 & -\half & 0 & 0 & 0 & 0 & \frac{\sqrt{3}}{2} \\
 0 & 0 & 0 & 0 & 0 & 0 & -\frac{\sqrt{3}}{2} & 0 \end{array}\right) \\
f_{7jk}=i\left(\begin{array}{cccccccc}
 0 & 0 & 0 & \half & 0 & 0 & 0 & 0 \\
 0 & 0 & 0 & 0 & \half & 0 & 0 & 0 \\
 0 & 0 & 0 & 0 & 0 & -\half & 0 & 0 \\
 -\half & 0 & 0 & 0 & 0 & 0 & 0 & 0 \\
 0 & -\half & 0 & 0 & 0 & 0 & 0 & 0 \\
 0 & 0 & \half & 0 & 0 & 0 & 0 & -\frac{\sqrt{3}}{2} \\
 0 & 0 & 0 & 0 & 0 & 0 & 0 & 0 \\
 0 & 0 & 0 & 0 & 0 & \frac{\sqrt{3}}{2} & 0 & 0 \end{array}\right) &
f_{8jk}=i\left(\begin{array}{cccccccc}
 0 & 0 & 0 & 0 & 0 & 0 & 0 & 0 \\
 0 & 0 & 0 & 0 & 0 & 0 & 0 & 0 \\
 0 & 0 & 0 & 0 & 0 & 0 & 0 & 0 \\
 0 & 0 & 0 & 0 & \frac{\sqrt{3}}{2} & 0 & 0 & 0 \\
 0 & 0 & 0 & -\frac{\sqrt{3}}{2} & 0 & 0 & 0 & 0 \\
 0 & 0 & 0 & 0 & 0 & 0 & \frac{\sqrt{3}}{2} & 0 \\
 0 & 0 & 0 & 0 & 0 & -\frac{\sqrt{3}}{2} & 0 & 0 \\
 0 & 0 & 0 & 0 & 0 & 0 & 0 & 0 \end{array}\right)
\end{array}
$$
These are clearly basis elements for the adjoint representation of the
algebra which is a subset of the SO(8) algebra.  

We now make the connection between the two by writing the basis elements
for SU(3) in terms of the SO(8) basis elements.  
\begin{eqnarray}
f_{1}&=&m_{23} +\half(m_{47}-m_{56}), \hspace{.8in}
  f_2 = -m_{13} +\half(m_{46}+m_{57}) \\
f_3 &=& m_{12}+\half(m_{45} - m_{67}),\hspace{.8in}
  f_4 = -\half(m_{17}+m_{26}+m_{35}) +\frac{\sqrt{3}}{2}m_{58} \\
f_5 &=& \half(m_{16}-m_{27}+m_{34}) -\frac{\sqrt{3}}{2}m_{48}, \;\;\;\; 
  f_6 = \half(-m_{15} + m_{24}+m_{37}) +\frac{\sqrt{3}}{2}m_{78} \\
f_7 &=& \half(m_{14}+ m_{25} -m_{36}) -\frac{\sqrt{3}}{2}m_{68}, \hspace{.2in}
  f_8 = \frac{\sqrt{3}}{2}(m_{45} + m_{67})
\end{eqnarray}
Let us now complete the basis by finding a complete set of 28 matrices
which are orthogonal in the Hilbert-Schmidt sense and have $f_i,\;
i=1,...,8$ as members.  Considering the Hilbert-Schmidt to be an inner
product, we can immediately write down a complete set.  We will number
them in no particular order.  (IMPORTANT:  We will also assume for now
that they are normalized such that Tr$(m_{ij}m_{kl})
=\delta_{ik}\delta_{jl}$.  This can be adjusted later if it is not
true.)  Let us consider a set of matrices which
span the space $m_{23},m_{47},m_{56}$ when $f_1$ is included.  The set
is (for $f_1$)
$$
f_9 = \half(m_{47} + m_{56}), \;\; f_{10} = -(m_{47} - m_{56}).
$$
Similarly, we will group the remaining matrices:
($f_2$)
$$
f_{11} = m_{13} + (m_{46} + m_{57}), \;\; f_{12} = (m_{46} -m_{57}),
$$
($f_3$ and $f_8$)
$$
f_{13} = m_{12}-(m_{45} - m_{67}),
$$
($f_4$)
$$
f_{14} = (m_{17}+m_{26} + m_{35}) +\sqrt{3}\; m_{58}, \;\; 
f_{15} = m_{17} -2m_{26} + m_{35}, \;\; 
f_{16} = m_{17} - m_{35}
$$
($f_5$)
$$
f_{17} = (m_{16} - m_{27} + m_{34}) + \sqrt{3} \; m_{48},\;\;
f_{18} = m_{16} +2m_{27} + m_{34}, \;\; 
f_{19} = m_{16} - m_{34}
$$
($f_6$)
$$
f_{20} = (-m_{15} + m_{24} + m_{37}) - \sqrt{3} \; m_{78},\;\;
f_{21} = m_{24} +2m_{15} + m_{37}, \;\; 
f_{22} = m_{24} - m_{37}
$$
($f_7$)
$$
f_{23} = (m_{14} + m_{25} - m_{36}) + \sqrt{3} \; m_{68},\;\;
f_{24} = m_{14} +2m_{36} + m_{25}, \;\; 
f_{25} = m_{14} - m_{25}.
$$
Finally, we have the following three:
$$
f_{26} = m_{18},\;\; f_{27} = m_{28}, \;\; f_{28} = m_{38}.
$$



\end{document}